\documentclass[11pt,a4paper]{article}

\usepackage{jcappub}
\usepackage[latin1]{inputenc}
\usepackage{bbm}
\usepackage{graphicx}
\usepackage{epsfig}
\usepackage{subfigure}
\usepackage{color}

\newcommand{\deldag}{\mathbin{\partial\mkern-10.5mu\big/}}

\newcommand{\be}{\begin{equation}} 
\newcommand{\ee}{\end{equation}}
\newcommand{\bea}{\begin{eqnarray}} 
\newcommand{\eea}{\end{eqnarray}}
\newcommand{\nn}{\nonumber}
\newcommand{\bmp}{\noindent\begin{minipage}{16cm}}
\newcommand{\emp}{\end{minipage}\vskip 7mm} 
\def\lsim{\mathrel{\raise.3ex\hbox{$<$\kern-.75em\lower1ex\hbox{$\sim$}}}}
\def\gsim{\mathrel{\raise.3ex\hbox{$>$\kern-.75em\lower1ex\hbox{$\sim$}}}}
\newcommand{\ie}{{\em i.e.~}}

\newcommand{\eg}{{\em e.g.~}}

\newcommand{\dsl}{\mathbin{\partial\mkern-10mu\big/}}
\newcommand{\intron}[1]{}

\newcommand{\N}{N}

\def\sfrac#1#2{{\textstyle\frac{#1}{#2}}}
\def\ww{{\rm \scriptscriptstyle W}}

\def\MNN{M_{\scriptscriptstyle \rm NN}}
\def\Mww{M_{\scriptscriptstyle ww}}
\def\Mbb{M_{\scriptscriptstyle \beta \beta}}
\def\mNb{m_{\scriptscriptstyle \rm N \beta}}
\def\mwb{m_{\scriptscriptstyle w \beta}}
\def\mNw{m_{\scriptscriptstyle {\rm N}w}}
\def\LNN{\lambda_{\scriptscriptstyle \rm NN}}
\def\Lww{\lambda_{\scriptscriptstyle ww}}
\def\Lwb{\lambda_{\scriptscriptstyle w\beta}}
\def\Lbb{\lambda_{\scriptscriptstyle \beta\beta}}

\def\mDM{m_{\scriptscriptstyle \rm DM}}
\def\ODM{\Omega_{\scriptscriptstyle \rm DM}}

\title{Dark matter from unification}

\author[a,b]{Kimmo Kainulainen} 
\author[a,b,d]{Kimmo Tuominen} 
\author[c]{Jussi Virkaj\"arvi}

\affiliation[a]{Department of Physics, University of Jyv\"askyl\"a, \\
                      P.O.Box 35 (YFL), FI-40014 University of Jyv\"askyl\"a, Finland}
\affiliation[b]{Helsinki Institute of Physics, \\
                       P.O.~Box 64, FI-00014 University of Helsinki, Finland}
\affiliation[c]{{CP}$^{ \bf 3}${-Origins} \& The Danish Institute for Advanced Study DIAS, \\
                University of Southern Denmark, Campusvej 55, DK-5230 Odense M, Denmark}
\affiliation[d]{Department of Physics and Astronomy, University of Southampton, 
                Southampton, SO17 1BJ, UK }
\emailAdd{kimmo.kainulainen@jyu.fi}
\emailAdd{kimmo.i.tuominen@jyu.fi}
\emailAdd{virkajarvi@cp3.dias.sdu.dk}

\abstract{We consider a minimal extension of the Standard Model (SM), which leads to unification of the SM coupling constants, breaks electroweak symmetry dynamically by a new strongly coupled sector and leads to novel dark matter candidates. In this model, the coupling constant unification requires the existence of electroweak triplet and doublet fermions singlet under QCD and new strong dynamics underlying the Higgs sector. Among these new matter fields and a new right handed neutrino, we consider the mass and mixing patterns of the neutral states. We argue for a symmetry stabilizing the lightest mass eigenstates of this sector and determine the resulting relic density. The results are constrained by available data from colliders and direct and indirect dark matter experiments. We find the model viable and outline briefly future research directions.}

\keywords{Dark matter, Naturality, Unification, Technicolor \\
\\
\emph{Preprint: CP3-Origins-2013-026  DNRF90,  Preprint: DIAS-2013-26}}

\begin{document}
\maketitle

%
\section{Introduction}
\label{Intro}

A new scalar boson, with properties compatible with the Standard Model (SM) Higgs boson, was discovered at the LHC experiments in July 2012~\cite{Aad:2012tfa,Chatrchyan:2012ufa}. This discovery, together with the determination of its properties with inclusion of more data~\cite{ATLAS-CONF-2013-012,CMS-PAS-HIG-13-001,ATLAS-CONF-2013-030,ATLAS-CONF-2013-013,ATLAS-CONF-2013-013,CMS-PAS-HIG-13-002,CMS-PAS-HIG-13-004,ATLAS-CONF-2012-161,CMS-PAS-HIG-12-044}, is providing stringent constraints on models of new physics beyond the Standard Model (BSM) 
\cite{Carmi:2012yp,Espinosa:2012ir,Giardino:2012ww,Alanne:2013dra}. Two currently much investigated model building paradigms are supersymmetry, in particular the minimal supersymmetric Standard Model (MSSM), and new strong dynamics (technicolor) sourcing the electroweak symmetry breaking (EWSB). For strong dynamics paradigm, the currently favored category of models is based on the idea of quasiconformality or walking \cite{Holdom:1984sk,Yamawaki:1985zg}, which means that that above the electroweak scale a new strongly coupled sector is governed by an approximate infrared fixed point of the new coupling. Concrete and currently viable realizations of this idea are minimal and next to minimal walking technicolor theories~\cite{Sannino:2004qp,Dietrich:2005jn}.\footnote{Note that a 125 GeV scalar particle is not {\em{generically}} a problem in technicolor: First, if the theory is near conformal, all mass scales are suppressed \cite{Dietrich:2005jn,Elander:2009pk,Grinstein:2011dq,Evans:2013vca}. Second, when the theory is coupled with the SM, there is further suppression from the electroweak interaction and most notably from the coupling of the Higgs-like scalar with the top quark \cite{Foadi:2012bb}.}

In addition to the electroweak sector of the SM, impetus for BSM model building is provided by the cosmological observation on the existence of the dark matter component in the energy content of the universe, as the SM does not provide a particle physics explanation for this observation. Currently several experiments, on Earth and onboard of satellites, are providing more data which constrains different models of dark matter. Both MSSM and technicolor provide dark matter candidates: In MSSM a dark matter candidate is the lightest supersymmetric particle which is stabilized by the postulated $R$-parity between ordinary particles and their superpartners. In technicolor on the other hand, the natural dark matter candidate is the lightest technibaryon, provided it is electrically neutral, as it is protected against decay by the technibaryon number (analogous to the ordinary baryon number).  

Technicolor provides the most elegant solution to the hierarchy problem, while MSSM provides for the unification of the SM coupling constants. In this paper we consider a minimal model setup which addresses both of these model building paradigms. The construction of the model has been discussed in detail in~\cite{Kainulainen:2010pk}. We review the model and extend the relevant features in section \ref{model} of the present paper. The main feature of the model is that as a consequence of addressing the naturalness and coupling constant unification, it also provides for a weakly interacting massive particle (WIMP) as a dark matter candidate. This WIMP is completely external to the strongly interacting technicolor sector. Of course also the lightest technibaryon, if electrically neutral, constitutes a candidate for dark matter in this model. However, in this paper we concentrate solely on the dark matter arising from the technicolor singlet fermion fields.
The dark matter analysis is described in section \ref{analysis} and the results and constraints are discussed in section \ref{constraints} of this paper. We find that in portions of the parameter space our WIMP candidate is compatible with current experimental constraints and able to explain all of the observed dark matter abundance. Moreover there naturally exists large portions of parameter space where our WIMP candidate provides for a subdominant dark matter component. 
Allowing also for technibaryonic dark matter in our model would make these scenarios interesting and pave way for further studies within this model.

%
\section{The model}
\label{model}
%

As mentioned in the introduction, our model is motivated by the possibility to combine different model building paradigms. We assume that the EWSB originates from a new strongly interacting sector (i.e. technicolor), and observe that inclusion of few additional degrees of freedom will force a very good one-loop unification of the SM coupling constants. Starting with such a setup we show that the model also features viable dark matter candidates.

The key observation~\cite{Gudnason:2006mk,Kainulainen:2010pk} is that 
one can obtain very good one-loop unification of the SM couplings if the SM matter spectrum is extended as follows: First, with respect to electroweak quantum numbers we add a full fourth generation of fermions. The lepton generation is denoted by $L_L$, $E_R$ and $\beta$. The ``quarks" $Q_L$, $U_R$ and $D_R$, however, are taken to be singlet under QCD, but are considered to transform in a three dimensional vector representation of a new SU($N_{\textrm{TC}}$) gauge group. Concretely this can be taken the fundamental representation of SU(3) or adjoint representation of SU(2). The latter is quasi-conformal with two flavors, while the former can be made quasi-conformal by adding further flavors in the fundamental representation of the new SU(3) but singlet under all SM charges. Second, then, we add one Weyl fermion, $\omega$, in the adjoint representation of SU(2)$_L$ and one adjoint Weyl fermion $\tilde{g}$ in the adjoint representation of SU(3)$_c$. The quantum number assignments for these new fields are shown in table \ref{chargeassignments}. The model is free of gauge and global anomalies. In this paper our goal is to provide a thorough analysis of a dark matter candidate emerging from this particle content.

\begin{table}[htb]
\begin{centering}
\begin{tabular}{|l||c|c|c|c|}
\hline
	& SU(3)$_c$ & SU(2)$_L$ & U(1)$_Y$ & SU($N_{\textrm{TC}}$) \\
	\hline
$L_L$ & 1 & 2 & -1/2 & 1 \\
$\beta$ & 1 & 1 & 0 & 1 \\
$E_R^c$ & 1 & 1 & 1 & 1 \\
$\omega$ & 1 & 3 & 0 & 1 \\
$Q_L$ & 1 & 2 & 1/6 & 3 \\
$U_R^c$ & 1 & 1 & 2/3 & 3 \\
$ D_R^c$ & 1 & 1 & 1/3 & 3 \\
$ \tilde{g}$ & adj. & 1 & 0 & 1\\
\hline
\end{tabular}
\caption{The table shows the new states added to SM, and their charge assignments under the SM gauge
group.}
\label{chargeassignments}
\end{centering}
 \end{table}

Concerning the mass spectrum of these states, we assume that the Weyl fermion, charged under QCD color, is very heavy and decoupled from low energy particle spectrum, but the one charged under SU(2) is relatively light. As the adjoint representation of SU(2) is three dimensional, the model contains a weak triplet state which includes one neutral particle. This neutral adjoint fermion along with the new heavy left handed neutrino contained in the doublet $L_L$ and the right handed singlet state $\beta$ are plausible dark matter particle candidates. Indeed, if we postulate that our new particles are invariant under a $Z_2$ symmetry, the lightest linear combination of these three neutral states becomes the naturally stable WIMP in our model. 

The technicolor sector is described by a low energy effective Lagrangian. For our purposes, this practically means that the new gauge theory sector is replaced with an effective composite Higgs doublet similar to the fundamental Higgs field of the SM. 
We consider that this effective SM-like Higgs field is the lightest new state in the effective model and sufficient to describe the low energy technicolor phenomenology needed in our DM studies. Thus we do not include other higher spin resonances, arising from the underlying strong dynamics, to our analysis. 

Several model building paradigms exists to address the origin of fermion masses and flavor patterns in the context of technicolor theories. The traditional possibility are the extended technicolor (ETC) interactions \cite{Dimopoulos:1979es,Eichten:1979ah,Appelquist:1993sg}. 
Alternative possibilities include the existence of additional fundamental scalar degrees of freedom mediating the electroweak symmetry breaking to the SM matter sector \cite{Simmons:1988fu,Kagan:1991gh,Carone:1992rh,Antola:2009wq} or supersymmetric technicolor 
\cite{Dine:1981za,Dobrescu:1995gz,Antola:2010nt,Antola:2011bx} where such scalar fields naturally arise.  Here we adapt a much simpler approach: We consider the effective low energy model including only the composite Higgs doublet, and take this as a convenient and gauge invariant means to parametrize the generation of the masses of chiral fermions and describe the resulting phenomenology. However, we have also introduced a singlet fermion, and in principle this can obtain its mass either from the Higgs field or from any other scalar present in the theory. As a simple extension of the scalar sector, we consider the case where the SM-like Higgs sector is extended with a single real EW singlet scalar $S$ which provides the mass for the singlet Weyl fermion $\beta$. 

The Lagrangian for the full low energy effective model can be schematically written as
\begin{equation}
{\mathcal{L}_{\rm{GM}}} = {\mathcal{L}_{4{\rm f}}} + {\mathcal{L}_{\rm{Ad}}} + {\mathcal{L}_{\rm{SM + Eff. Higgs}}} 
                     + {\mathcal{L}_{\rm{rest}}}  \,,
\label{eq:Ltot}
\end{equation}
where ${\mathcal{L}_{4\rm{f}}}$ introduces the $4^{\rm{th}}$ lepton family, ${\mathcal{L}_{\rm{Ad}}}$ the SU(2) adjoint Weyl fermions including also the right handed singlet state, ${\mathcal{L}_{\rm{SM+ Eff. Higgs}}}$ the SM without fundamental Higgs field but with the low energy effective Higgs doublet, and finally ${\mathcal{L}_{\rm{rest}}}$ denotes rest of the TC sector, SU(3) adjoint Weyl fermions and possible other terms which are not important from the DM analysis viewpoint. 

Now we turn our attention to the parts of the model Eq.~(\ref{eq:Ltot}), which are relevant for the DM studies. In particular we will write explicitly the gauge interactions for the new leptons and for the SU(2) adjoint fermions. Then we will consider their interactions with the effective Higgs field leading to the mass and mixing patterns, and we determine the DM-gauge and DM-Higgs interactions relevant for the computation of the DM relic abundance. 

%
\subsection{Weak currents}
%

\emph{Weak currents for heavy leptons}. We denote the left handed heavy lepton doublet with $L_L = (N_L E_L)^T$ and the charged right handed singlet with $E_R$. Using the SM-like hypercharge assignments the kinetic and gauge interaction terms for the new leptons become equal to the corresponding terms of SM leptons. The Lagrangian for heavy leptons in the charge eigenbasis thus reads 
\begin{eqnarray}
{\mathcal{L}_{4.\rm{f}}}  
\,=\,  i\bar{L}_L \dsl L_L \,+\, i\bar{E}_R \dsl E_R \,+\, \mathcal{L}_{\rm{4f,H}} 
\, +\, {\mathcal{L}_{\rm{W}}} \,+\, {\mathcal{L}_{\rm{Z}}} \,+\, {\mathcal{L}_{\rm{EM}}} \,,
\label{eq:Llep}
\end{eqnarray}
where the mass-Lagrangian $\mathcal{L}_{\rm{4f,H}}$ will be introduced in Sec.~\ref{sec:massterms} below. The weak and electromagnetic currents are
\begin{eqnarray}
{\mathcal{L}_{\rm{W}}} & = & \frac{g}{\sqrt{2}}\left(W_{\mu}^{-} \bar{E}_L \gamma^{\mu} N_L + W_{\mu}^{+}\bar{N}_L \gamma^{\mu} E_L\right), \nn \\ 
 {\mathcal{L}_{\rm{Z}}} & = & \frac{g}{2c_\ww} Z_{\mu} 
 \Big( \bar{N}_L \gamma^{\mu} N_L + ( 2s_\ww^2 -1) \bar{E}_L \gamma^{\mu} E_L
  + 2s_\ww^2\bar{E}_R \gamma^{\mu}E_R \Big) \,, \nn \\
{\mathcal{L}_{\rm{EM}}} & = & -eA_{\mu} \bar{E} \gamma^{\mu}E \,,\phantom{\frac{1}{1}}
\label{eq:Lcur}
\end{eqnarray} 
where $g$ is the weak coupling constant and $c_\ww \equiv \cos\theta_W$ and $s_\ww \equiv \sin\theta_W$, where $\theta_W$ is the Weinberg angle.
\\ 

\noindent \emph{Weak currents for the SU(2) adjoint Weyl fermions}. The Lagrangian density for the left handed SU(2) adjoint Weyl fermions and for the right handed singlet Weyl fermion reads  
\begin{equation}
{\mathcal{L}_{\rm{Ad}}} =  i\omega^{\dagger}\bar{\sigma}^\mu D_\mu \omega 
               + i\beta\sigma^\mu\partial_\mu \beta^{\dagger} + {\mathcal{L}_{\rm{Ad,H}}} \,.
\label{eq:Lad}
\end{equation} 
Here $\omega = (w^1,w^3,w^3)$ denotes the left-handed SU(2) weak triplet and $\beta^{\dagger}$ the right-handed singlet state.  We use the usual notations
$\bar{\sigma}^\mu \equiv (1, -\vec{\sigma})$ and $\sigma^\mu \equiv (1, \vec{\sigma})$ where the components of $\vec{\sigma}$ are the Pauli matrices. Finally, the covariant derivative is, in component form,
\begin{equation}
D^{ac}_\mu =  \partial_\mu \delta^{ac} + g \epsilon^{abc} A^b_\mu \,,
\label{eq:covder}
\end{equation}
where the SU$_L$(2) generators in the adjoint representation are denoted with $[T^a]_{bc}= -i\epsilon^{abc}$. 
The triplet $\omega$ is charged only under weak isospin and the field $\beta^\dagger$ is a pure singlet state, not charged under any gauge groups. 

Now we want to write these 2-component Weyl fields as 4-component Dirac and Majorana fields which will be more practical to use when performing the matrix element calculations for the annihilation processes. To this end we need to first find the eigenstates of the charge operator $Q=T^3+Y$. The field $\omega$ is a hypercharge singlet, hence the eigenstates of the diagonal weak isospin operator $T^3$ are directly the charge eigenstates which we are looking for:
\begin{eqnarray}
w^\pm=\frac{1}{\sqrt{2}}(w^1\mp iw^2), \qquad w^0  =w^3 \,,
\end{eqnarray}
where the superscripts $\pm 1$ and 0 refer to the electromagnetic charges. In this basis the kinetic term becomes
\begin{equation}
i \omega^{\dagger}\bar{\sigma}^\mu\partial_\mu \omega = 
 iw^{+\dagger}\bar{\sigma}^\mu\partial_\mu w^+
+iw^{-\dagger}\bar{\sigma}^\mu\partial_\mu w^-
+iw^{0\dagger}\bar{\sigma}^\mu\partial_\mu w^0 \,,
\label{free_L}
\end{equation} 
and the gauge interaction term reads 
\begin{eqnarray}
{\mathcal{L}}_{\rm gauge} & = & ig \epsilon^{abc} w^{a\dagger} \bar{\sigma}^\mu A_\mu^b w^c \\  
		& = &	g\left( \begin{array}{ccc} w^{+ \dagger} w^{0 \dagger} w^{- \dagger} \end{array} \right) \bar{\sigma}^\mu
				\left( \begin{array}{ccc} W_{\mu}^{3} & -W_{\mu}^{+} & 0     \\
				                         -W_{\mu}^{-} &       0      & W_{\mu}^{+} \\
				                                   0  & W_{\mu}^{-}  & -W_{\mu}^{3} \end{array} \right)
				\left( \begin{array}{ccc} w^{+} \\ w^{0} \\ w^{-}  \end{array} \right)  \,,
\label{eq:gaugeint1}
\end{eqnarray}
where $W_\mu^\pm = (A^1_\mu \mp i A^2_\mu)/\sqrt{2}$ are the usual charged gauge bosons. Defining 4-component Dirac spinors carrying negative and positive charges,
\begin{equation}
w_D^{-}=\left(\begin{array}{c} {w}^-_\alpha \\ (w^+)^{\dagger\dot{\alpha}}\end{array}\right),~~
w_D^{+}=\left(\begin{array}{c} {w}^+_\alpha \\ (w^-)^{\dagger\dot{\alpha}}\end{array}\right) \, ,
\end{equation}
and the 4-component neutral Majorana spinors
\begin{equation}
w_M^0=\left(\begin{array}{c} {w}^0_\alpha \\ 
(w^0)^{\dagger \dot{\alpha}}\end{array}\right), ~~
\beta_M=\left(\begin{array}{c} \beta_\alpha \\ 
\beta^{\dagger\dot{\alpha}}\end{array}\right) \,,
\label{eq:neutral}
\end{equation}
all previous expressions written in 2-component notation can be transformed to a 4-component form.
Since the Dirac and Majorana fields satisfy relations: $(w_D^{-})^{c}=w_D^+$ and $(w_M^{0})^{c}=w_M^0$ and $(\beta_M)^{c}=\beta_M$, everything can be written using only $w_M^0$, $\beta_M$  and either of the charged Dirac-spinors, say $w_D \equiv w_D^-$. The Lagrangian (\ref{eq:Lad}) is
\begin{eqnarray}
{\mathcal{L}}_{\rm{Ad}}
&=& i\bar w_D \deldag w_D 
  + i\bar w_M \deldag w_M 
  + i\bar \beta_M \deldag\beta_M + {\mathcal{L}_{\rm{Ad,H}}}
\nonumber \\
&+& g\left(
    W_\mu^+\bar{w}^0_M\gamma^\mu w_D
   +W_\mu^-\bar{w}_D\gamma^\mu w^0_M
   -W^3_\mu\bar{w}_D\gamma^\mu w_D \right) \, ,
\label{gauge_lagrangian}
\end{eqnarray}
providing the charged and neutral currents:
\begin{eqnarray}
\label{ew_currents}
{\mathcal{L}}_W &=& g\,(W_\mu^+\bar{w}^0_M\gamma^\mu w_D+W_\mu^-\bar{w}_D\gamma^\mu w^0_M),\\
{\mathcal{L}}_Z &=& g c_\ww\, Z_{\mu}\bar{w}_D\gamma^\mu w_D, \nonumber \\
{\mathcal{L}}_A &=& e\, A_{\mu}\bar{w}_D\gamma^\mu w_D \nonumber \, .
\end{eqnarray}
Electroweak gauge fields do not couple directly to the singlet field $\beta_M$. However, an effective coupling for a `bino'-like WIMP will be induced by mass mixing, which is considered in Eqs.~(\ref{4fhiggsmass}) and (\ref{Adhiggsmass}) below. Further, from Eq.~(\ref{ew_currents}) we see that $w_M^0$ does not couple to the neutral $Z$ boson. This can have significant impact for the DM analysis. Indeed, if the WIMP is dominantly a mixture of $w_M^0$ and $\beta_M$, then its relic density is determined by the effective Higgs boson interactions and by the charged current processes, and only spin independent interactions are relevant for its direct detection. On the other hand, if the WIMP is dominantly a mixture of $N_L$ and $\beta_M$ it always couples to $Z$. In this case its relic density may be substantially affected by $Z$ boson interactions, and also spin dependent interactions are relevant for its direct detection.

%
\subsection{Effective mass terms}  
\label{sec:massterms}
%

We will now introduce an effective Higgs doublet $H$ into the low energy realization of our  
model. In the dynamical EW symmetry breaking this will generate the mass terms and effective Higgs couplings for the fourth family leptons, SU(2) adjoint particles and for the EW singlet. First, we introduce interaction terms between $H$, the $4^{\rm{th}}$ family heavy leptons and the neutral singlet field $\beta_M$. The gauge invariant effective interactions, up to dimension five operators, which we include, are
\begin{equation}
{\mathcal{L}}_{\rm{4f,H}} 
  = y_E \bar{L}_L H E_R 
  + y_\beta\bar{L}_L\tilde{H}\beta_{R} 
  + \frac{\LNN}{\Lambda}(\bar{L}^c\tilde{H})(\tilde{H}^TL)+{\rm{h.c.}} \,,
\label{4fhiggsmass}
\end{equation}
where $\tilde{H}=i\tau^2H^\ast$, and $y_E$, $y_\beta$ and $\LNN$ are some dimensionless coupling constants. The first two terms in Eq.~(\ref{4fhiggsmass}) are the usual SM-like Yukawa couplings. The first one generates Dirac mass terms for the charged lepton $E$ and the second for a Dirac neutrino, whose right handed component is the sterile state $\beta$. The last term in Eq.~(\ref{4fhiggsmass}) is a non-renormalizable dimension five operator, which produces a left handed Majorana mass for the neutrino $N$.

Note that the interactions in Eq.~(\ref{4fhiggsmass}), as well as the other terms in Eqs.~(\ref{eq:Llep}) and (\ref{gauge_lagrangian}), are invariant under $Z_2$ symmetry transformation, in which $E\rightarrow -E$, $N \rightarrow -N$, $\beta \rightarrow -\beta$ and $w\rightarrow -w$. On the other hand, gauge invariance would allow to write following Yukawa couplings:
\begin{eqnarray}
& & y_{wi} H^T(i\tau^2)\omega L_i+y_{\beta i}H^T(i\tau^2)\beta L_i + {\rm{h.c.}} \nonumber \\
&=& -\frac{v+h}{\sqrt{2}}[y_{wi}(\sqrt{2}w^+e_i+w^0\nu_i)+y_{\beta i}\beta\nu_i] + {\rm{h.c.}} \,,
\label{danger_zone}
\end{eqnarray}
where the $L_i$ is generation $i$ SM lepton doublet. These terms couple the SM fields to the new adjoint and singlet fields and so a WIMP containing a non-negligible mixture of these states would become unstable against decay into light SM particles.  However, if SM field are singlets under the $Z_2$ symmetry introduced above, then these interactions are forbidden and the stability of the DM candidate is guaranteed by a symmetry principle. We note that this symmetry, as well as the non-renormalizable interactions in Eq.\,(\ref{4fhiggsmass}), may originate from a UV complete theory responsible for the generation of the full flavor structure of the model above the scale $\Lambda\gg v_{w}$, but we shall not elaborate on such details here.

After the spontaneous symmetry breaking (SSB) where $H\rightarrow (0,v+h)^T /\sqrt{2}$, Eq.~(\ref{4fhiggsmass}) becomes 
\begin{eqnarray}
{\mathcal{L}}_{\rm{4f,H}}
\; &\stackrel{{\rm{SSB}}}{\rightarrow}& \;
           m_E \bar{E} E  \, \Big(1 + \frac{h}{v} \Big) \nn \\
        &+&\frac{\mNb}{2} (\overline {\beta}_R N_L  + \overline {\beta}_L N_R  ) \, 
                           \Big(1 + \frac{h}{v} \Big) + {\rm{h.c.}} \nn \\
        &+&\frac{\MNN}{2}{\bar N}_M N_M \, \Big(1 + \frac{h}{v} \Big)^2 \,, 
\label{4fhiggsmass2}
\end{eqnarray}
where the masses are $m_E \equiv y_E v/\sqrt{2}$, $\mNb \equiv y_\beta v/\sqrt{2}$ and $\MNN \equiv \LNN v^2/\Lambda$, and $v$ is the vacuum expectation value (VEV) of the composite Higgs field $h$. Note that $(N_{M})^c = N_M$. Here and in the following the shorthand 4-component notations $N_L \equiv N_{ML} = (N_\alpha,0)^T$, $N_R \equiv N_{MR} = (0, N^{\dagger\dot \alpha})^T$, $w^0_L \equiv w^0_{ML} = (w^0_\alpha,0)^T$ and $\beta_R \equiv \beta_{MR} = (0,\beta^{\dagger\dot \alpha})^T$ are used.

Next we consider the Higgs couplings to the SU(2) adjoint fields, consistent with the $Z_2$-symmetry, again including operators up to dimension five:
\begin{equation}
{\mathcal{L}}_{\rm{Ad,H}}
 =  y_w \tilde{H}^T\omega L_L 
  + \frac{\Lwb}{\Lambda} \beta H^\dagger\omega H 
  + \frac{\Lww}{\Lambda} H^\dagger\omega\omega H \, + \, {\rm{h.c.}}  \,,
\label{Adhiggsmass} 
\end{equation}
where $\omega\equiv \omega^a\tau^a$ and $\tau^a=\sigma^a/2$ in terms of the Pauli matrices. The scale $\Lambda$ is the same we introduced in Eq.~(\ref{4fhiggsmass}). When $\sqrt{2}H\rightarrow (0,v+h)^T$, Eq.~(\ref{Adhiggsmass}), written in the 4-component notation, becomes
\begin{eqnarray}
{\mathcal{L}}_{\rm{Ad,H}} 
   \; &\stackrel{{\rm{SSB}}}{\rightarrow}&  \; 
 \frac{\mNw}{2} (\bar {w}^0_R N_L + \bar {w}^0_L N_R ) \, 
   \Big(1 + \frac{h}{v} \Big) + {\rm{h.c.}} 
   \nn \\
        &+&\frac{\mwb}{2} (\bar {\beta}_R w^0_L  + \bar {\beta}_L w^0_R  ) \, \Big(1 + \frac{h}{v} \Big)^2 + {\rm{h.c.}}
   \nn \\
        &+&( \Mww \bar{w}_D w_D + \frac{\Mww }{2}\bar{w}_M^0w_M^0 ) \,
 \Big(1 + \frac{h}{v} \Big)^2\,.
\label{Adhiggsmass2}
\end{eqnarray}
Here $\Mww \equiv \Lww v^2/4\Lambda$ is both the mass of the charged adjoint field $w_D$ and the Majorana mass of the neutral adjoint state $w_M^0$. The two Dirac mixing masses are defined as $\mNw \equiv  y_w v/2\sqrt{2}$ and $\mwb \equiv \Lwb v^2/2\Lambda$.  

Finally, we can provide a Majorana mass for the singlet field $\beta_M$ either via a dimension five interaction with the Higgs, $(\Lbb/\Lambda) H^\dagger\beta\beta H$, or through a VEV of a new weak SU(2) singlet field $S$, which can plausibly emerge from a more complete extended technicolor theory. In the former case the WIMP is very strongly coupled to Higgs, and hence heavily constrained by the direct DM searches~\cite{Heikinheimo:2012yd}. We therefore choose the latter option and include the following gauge- and $Z_2$ symmetric interaction Lagrangian: 
\begin{eqnarray}
{\mathcal{L}}_{\beta S} = y_R S \beta\beta + {\rm{h.c.}} \; \stackrel{SSB}{\rightarrow} \;
 \frac{\Mbb}{2}\overline{\beta}_M\beta_M \Big(1 + \frac{s}{v_s} \Big) \,.
\label{singlet} 
\end{eqnarray}
The right handed Majorana mass now is $M_R \equiv \sqrt{2}y_Rv_s$ where $v_s$ is the VEV of the singlet field $S$. 

%
\subsection{Mass mixing and couplings}
\label{sec:massmixing}
%

The general $3\times3$ mass matrix of the new neutral Majorana particles $N_M,w^0_M$ and $\beta_M$ can now be formed by collecting all mass terms from the equations (\ref{4fhiggsmass2}), (\ref{Adhiggsmass2}) and (\ref{singlet}). The resulting matrix in the four component notation reads
\begin{eqnarray}
{\mathcal{L}}_{\rm{mass}} 
  = \frac{1}{2} \left( \begin{array}{ccc} \overline{N}_R, \, \overline{w^0}_R,  \, \overline{\beta}_R \end{array} \right) 
				\left( \begin{array}{ccc} \MNN & \mNw  & \mNb \\
				                          \mNw & \Mww  & \mwb   \\
				                          \mNb & \mwb  & \Mbb \end{array} \right)
				\left( \begin{array}{ccc} N_L \\ w^0_L \\ \beta_L  \end{array} \right)  + {\rm{h.c.}}
\label{eq:massmatrix}
\end{eqnarray}
This mass matrix induces a mixing pattern analogous to that described by the Pontecorvo-Maki-Nakagawa-Sakata (PMNS) matrix in the usual $3 \times 3$ light neutrino mixing. Note that here the mixing matrix appearing in charged weak currents is actually a $2 \times 3$-matrix, since there are only two charged Dirac fields (in contrast with the three charged leptons present in the usual neutrino mixing case), coupled to our three neutral fields. However the mass matrix diagonalizing proceeds in a similar way as in the case of $3 \times 3$ light neutrino mixing. We shall not use the PMNS parameterization for the mixing matrix here. Instead, we use the Dirac and Majorana masses as the primary parameters and diagonalize the mass matrix numerically to obtain the weak currents and the effective Higgs interactions in the mass eigenbasis. Indeed, the symmetric mass matrix $M$ appearing in Eq.~(\ref{eq:massmatrix}) can be diagonalized with a unitary transformation $U^{T} M U = m$, where the mass eigenvalues are $m_i\geq 0$ and $U$ is an unitary matrix. Using the notation $\Omega_L \equiv (N_L ,\, w^0_L, \, \beta_L )^T$ and the relation $U^{\dagger} U = 1$, Eq.~(\ref{eq:massmatrix}) can be written in the form
\begin{equation}
{\mathcal{L}}_{\rm{mass}} \; = \; \frac{1}{2} \overline{\Omega}_R M \Omega_L  
                             +\frac{1}{2} \overline{\Omega}_L M^{\dagger} \Omega_R                            \; = \; \frac{1}{2} \overline{\chi} m \chi \,.
\label{eq:masseigenvalues}
\end{equation}
Here $m$ is the diagonal mass matrix with positive mass eigenvalues. The corresponding mass eigenstates are given by 
\begin{eqnarray}
\chi = \chi_L + \chi_R \equiv U^{\dagger} \Omega_L +  U^{T} \Omega_R \,,
\label{eq:masseigenstates}
\end{eqnarray}
and they obey the Majorana condition $\chi^c_i = \chi_i$. Eq.~(\ref{eq:masseigenstates}) can be immediately inverted
\begin{eqnarray}
\Omega_L = U \chi_L \,,  \quad 
\Omega_R = U^* \chi_R \, .
\label{eq:staterelations}
\end{eqnarray}

Using Eqs.~(\ref{eq:Lcur}), (\ref{ew_currents}) and (\ref{eq:staterelations}) we can write the weak currents of the heavy leptons and of the SU(2) adjoint fermions in the mass eigenbasis in terms of the mass eigenfields $\chi_i$:
\vskip0.01truecm
\begin{eqnarray}
{\mathcal{L}^{\rm{W}}_{\rm{4f}}} 
& = & \frac{g}{\sqrt{2}} W_{\mu}^{-} 
\sum_i U_{1i} \bar{E}_L \gamma^{\mu} \chi_{iL}  + \rm{h.c.}  \, , 
\label{Wcur} \\ 
{\mathcal{L}^{\rm{Z}}_{\rm{4f}}} 
& = & \frac{g}{2c_\ww} Z_{\mu}
\Big(\sum_i |U_{1i}|^2 \, \bar{\chi}_{iL} \gamma^{\mu} \chi_{iL} 
+ \sum_{i>j} \bar{\chi}_{i}(iV_{ij}+A_{ij}\gamma^5)\gamma^{\mu} \chi_{j} \Big) \,,
\label{Zcur} \\
{\mathcal{L}}^{\rm{W}}_{\rm{Ad}} 
& = & g W_\mu^- \sum_i  \overline{w}_D (V_i+iA_i\gamma^5)\gamma^\mu\chi_{i}  + \rm{h.c.} \,,
\label{eq:weakcur-mbasis}
\end{eqnarray}
\vskip-0.3truecm
\noindent where
\begin{equation}
\label{AB}
V_{ij} = {\Im}(U^*_{1i} U_{1j}),  \quad
A_{ij} \equiv \Re(U^*_{1i} U_{1j}), \quad
V_{i} \equiv  \Re (U_{2i})  \;\; {\rm and} \;\; 
A_{i} \equiv \Im(U_{2i}) 
\end{equation}
and $U_{ij}$ are the numerically determined elements of the diagonalizing matrix $U$. Further, using Eqs.~(\ref{4fhiggsmass2}-\ref{Adhiggsmass}), (\ref{singlet}) and (\ref{eq:staterelations}) we find that the Higgs interactions in the mass eigenbasis become:
\begin{equation}
{\mathcal{L}_{h \chi}} = 
 -\frac{gh}{2m_W}  \sum_{i\le j}  
          \bar{\chi}_{i} (S_{ij}  + P_{ij}  \gamma^5)\chi_{j} 
     -\frac{g^2h^2}{4m_W^2}  \sum_{i} \bar{\chi}_{i} (S^{2}_{ii} + P^{2}_{ii} \gamma^5) \chi_{i}
+ \ldots \,,
\label{eq:Higgscoup-mbasis}
\end{equation} 
where dots refer to additional terms which do not affect the tree level matrix element calculations. Here $m_W$ is $W^{\pm}$-boson mass. The mixing angle and mass dependent coefficients 
are defined as
\begin{eqnarray}
S_{ij}  &=&  - \mNb A_{ij} + (\delta_{ij}-2) \Mbb D_{ij}+ m_i\delta_{ij}  \,,
\nonumber \\
P_{ij} &=&  - \mNb i V_{ij} - i(\delta_{ij}-2) \Mbb E_{ij}  \,,
\nonumber \\
S^{2}_{ii} &=& - \mNb A_{ii} - \sfrac{1}{2} \Mbb D_{ii} + \sfrac{1}{2} m_i  \,,
\nonumber \\
P^{2}_{ii}&=& - \mNb i V_{ii} +\sfrac{i}{2} \Mbb E_{ii}  \,,
\label{eq_cfactors}
\end{eqnarray}
where $m_i$ is the i'th mass eigenvalue, the projection factors $V_{ij}$ and $A_{ij}$ are as defined in Eq.~(\ref{AB}) and 
\begin{equation}
D_{ij} \equiv \Re(U_{3i} U_{3j}) \quad {\rm and} \quad 
E_{ij} \equiv \Im(U_{3i} U_{3j}) \,.
\label{DE}
\end{equation}
Equations (\ref{Wcur}-\ref{DE}) contain all information needed to calculate the WIMP annihilation matrix elements needed in the DM freeze-out calculation in section \ref{sec:relic}.  

%
\section{Model analysis}
\label{analysis}
%

At low energies the parameter space of our model is entirely spanned by the three Majorana masses $M_{ii}$ and the three Dirac masses $m_{ij}$, with $i,j=N,\beta,w$, entering the $3\times 3$ mixing matrix of neutral states in Eq.~(\ref{eq:massmatrix}), and by the Dirac mass $m_E$ of the new charged state $E$. For us it is more sensible to use these Lagrangian parameters, rather than the physical masses and mixings, as input. Since Lagrangian masses can be directly linked to the effective couplings $y_E$, $y_\beta$, $y_w$, $y_R$, $\LNN$, $\Lww$, $\Lwb$ and the scales $v$, $v_s$ and $\Lambda$, we can more easily infer reasonable prior ranges for them. Naive dimensional analysis implies that $y_i < 4\pi$ and $\lambda_{ij} < (4\pi)^2$. Then, assuming $\Lambda$ to be in the TeV scale, we find it reasonable to adopt the following ranges:
\begin{equation}
| M_{ij} | \leq 3000 \;{\rm{GeV}}; \quad
 | m_{ij}  |  \leq 2000 \;{\rm{GeV}} \quad {\rm and} \quad  200\,{\rm{GeV}} \leq m_E \leq 2000\;{\rm{GeV}} \,,
\label{eq:pmrange}
\end{equation}
although our results are not particularly sensitive to the precise values of the upper limits for these masses. We scan the parameter range defined in Eq.~(\ref{eq:pmrange}) using Monte Carlo Markov Chain (MCMC) methods to find parameter values that pass all the experimental and observational constraints.

Given a random realization of the seven mass parameters we first diagonalize the neutral DM mass matrix numerically to find the mass eigenvalues $m_i$ and the diagonalizing matrix elements $U_{ij}$. From these we identify the lightest neutral eigenstate as the WIMP and construct the necessary WIMP-Higgs and the WIMP-gauge boson couplings. We then subject the parameter set to the oblique electroweak parameter test and the $Z$-boson invisible decay width constraint. If the set passes these constraints, we solve the relic density using a fast but accurate approximative method~\cite{Cline:2013gha}, and compare it with the DM density range that is consistent with the most recent observational data~\cite{Ade:2013zuv}. If the set passes also this test, we successively check it against the constraints on invisible Higgs branching ratio, spin independent and spin dependent XENON100 cross section limits and similar IceCube, COUPP and Super Kamiokande constraints. If the parameter set under consideration passes also these tests, we re-calculate the relic density using accurate numerical methods and save the parameter set and the results.  

After each successful step, a new Monte Carlo step is taken by randomly varying parameters around the successful solution using gaussian distribution with variance 
$(\sigma_{\rm R,I}/{\rm GeV})^2 = 2000$ for complex valued variables, and $(\sigma/{\rm GeV})^2 = 2000$ for real valued variables within the prior bound-areas. However, if at some point during the test chain the parameter set fails, a Monte Carlo Markov Chain step is taken: with a 15 $\%$ probability this unsuccessful set is still selected as a new starting point and otherwise the previous set is taken as a starting point for the next Monte Carlo step. This procedure is run until the whole prior restricted parameter space is scanned effectively. Because of the high dimensionality of the parameter space, and because many of the constraints are fulfilled only in small and often orthogonal subspaces of parameters, finding acceptable points takes very long MCMC calculation chains.

%
\subsection{Relic density}
\label{sec:relic}
%

The relic density of thermal WIMPs can be solved from the Lee - Weinberg equation~\cite{Lee:1977ua,Kainulainen:2009rb}:
\begin{eqnarray}
\frac{\partial Y_\chi}{\partial x} 
        = \frac{\langle v\sigma\rangle 
         m^3 x^2}{H} (Y_\chi^2-Y_{\rm eq}^2) \,.  
\label{ecosmo1}
\end{eqnarray}
Here $Y_\chi \equiv n_{\chi}/{s_E}$ is the ratio of the WIMP number density $n_{\chi}$ and the entropy density $s_E$. The integration variable encoding the adiabatic expansion law is $x \equiv s_E^{\scriptscriptstyle 1/3}/m$ and the Hubble parameter $H = (8\pi\rho/3M_{\rm Pl}^2)^{1/2}$, where $\rho$ is the energy density of the universe. We are assuming a standard expansion history of the universe, so that $\rho$ and $s_E$ are fully set by the particle content of the standard model.  The thermally averaged annihilation cross section $\langle v\sigma \rangle$ is~\cite{Gondolo:1990dk}:
\begin{equation}
   \langle v \sigma \rangle = 
       \frac{1}{8m_i^{4}TK^{2}_2(\frac{m_i}{T})}
      \int_{4m_i^2}^{\infty}ds
                        \sqrt{s}(s-4m_i^2)K_1(\frac{\sqrt{s}}{T}) 
                        \sigma_{\rm tot}(s) \,,
\label{ecosmo2}
\end{equation}
where $s$ is the Mandelstam invariant and $K_i(y)$, with $i=1,2$, is the modified Bessel function of the second kind. As was first observed in~\cite{Enqvist:1988we}, it is important to include all gauge- and Higgs boson final states in the total annihilation cross section $\sigma_{\rm tot}$ of heavy WIMPs. Here we consider all the relevant processes: $\chi_{i}\overline{\chi}_{i} \rightarrow$ $f\bar{f}$, $WW$, $ZZ$, $Zh$ and $hh$, where $f$ are the SM fermions, $W^{\pm}$ and $Z$ are the SM weak gauge bosons and $h$ is the effective composite Higgs boson. We computed and used complete cross sections to all these final states without any approximations. However the resulting expressions are too long to be reproduced here. We do not include annihilation channels to techniquarks or to possible low energy techni-resonances or the singlet $S$ giving rise to the singlet mass term in Eq.~(\ref{singlet}). This is justifiable if these states are very heavy. Indeed if these resonances have masses of order $\Lambda_{\rm TC} \sim $ 1 TeV, they could be relevant only for very heavy WIMPs, but even then the annihilations to (techni)fermionic channels would be sub-leading in comparison to the dominant $W$ gauge boson channels. Given the value of $Y_\chi(0)$ today, we obtain the WIMP relic density  $\Omega_\chi h^2$ from
\begin{eqnarray}
\Omega_\chi h^2 \simeq 2.7 \times 10^8 \left( \frac{m_i}{\rm GeV} \right) Y_\chi(0) \,.
\label{ecosmo3}
\end{eqnarray}
Computing $Y_\chi(0)$ numerically from Eq.~(\ref{ecosmo1}) is straightforward, but relatively time consuming. To speed up the search algorithm we at first use a much more efficient, yet accurate, freeze-out formalism described for example in refs.~\cite{Enqvist:1988we,Enqvist:1988dt} (for recent improvements see refs.~\cite{Steigman:2012nb,Cline:2013gha}). However, for all parameter values in our final sample, we recompute $Y_\chi(0)$ by directly solving equation (\ref{ecosmo1}) numerically.

%
\section{Constraints and results}
\label{constraints}
%

In this section we will present our results and describe in detail the relevant experimental and observational constraints which we apply on our model in the MCMC analysis. The very first constraint we apply for all parameter sets is that the lightest neutral particle is stable; that is the lightest one of all non-singlet states under the $Z_2$ symmetry, including the charged particles $E$ and $w_D$. Other constraints are due to electroweak precision measurements, direct dark matter searches, direct searches at colliders and indirect dark matter detection. We now proceed to discuss these in detail.

%
\subsection{Oblique constraints}
%

\begin{figure}
\begin{center}
\includegraphics[width=0.7\textwidth]{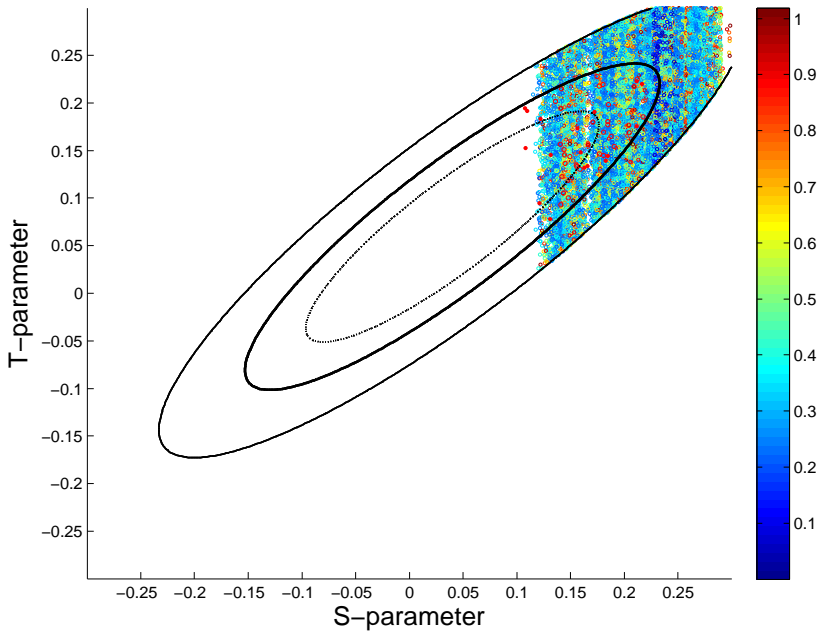}
\caption{The points are projections of parameter sets plotted in $S$ and $T$ plane. Coloring of data points encodes the relic density parameter $f_{\rm rel}$, defined in Eq.~(\ref{eq:frel}), as indicated by the vertical bar on the right. Ellipses depict the experimental 1$\sigma$, 1.6$\sigma$ and 2.6$\sigma$ allowed contours. The shown data points also pass the invisible Higgs width constraint $R_{\rm I} < 0.28$ and the spin independent XENON100 2011 limits. }
\label{fig:STplot}
\end{center}
\end{figure}

Let us start with the oblique corrections. All particles that carry electroweak quantum numbers contribute to the electroweak gauge boson polarization amplitudes. These amplitudes are usually characterized in terms of the oblique parameters $S$ and $T$~\cite{Peskin:1990zt}, whose values are severely constrained by the EW precision data. The new doublet and adjoint SU(2) states in our model, as well as the new states in the TC sector are charged under SU(2), whereby we need to test our model against the EW precision data. Our analysis is similar to the one of ref.~\cite{Antipin:2009ks}, which considered a similar TC-model, but with only one additional heavy SU(2)-doublet. In comparison to ref.~\cite{Antipin:2009ks} we have new contributions to $S$ and $T$ coming from the charged Dirac field $\omega^{\pm}_D$, and from three mixing neutral states instead of a two neutral particles. Explicit expressions for the $S$ and $T$ parameters are given in the appendix~\ref{sec:appendix}. We use the experimental constraints:
\begin{equation}
{S} = 0.04 \pm 0.09,  \quad {\rm and} \quad {T} = 0.07 \pm 0.08  \,,
\label{expST}
\end{equation}
which include the 88$\%$ correlation between $S$ and $T$ as given by~\cite{Beringer:1900zz}\footnote{The  quoted values assumed $115.5$ GeV$<m_H<127$ GeV and is therefore consistent with the observed $m_H=125$ GeV.}. In figure \ref{fig:STplot} we have plotted the parameter space points  falling within the 1$\sigma$, 1.6$\sigma$ and 2$\sigma$ constraint ellipses in the $(S,T)$-plane and passing the invisible Higgs width constraint $R_{\rm I} < 0.28$ and the spin independent XENON100 2011 limits. These various constraints will be described in more detail in the following subsections. The color coding of the points expresses the value of the fraction of the observed relic density for each given parameter set as indicated by the bar to the right of the figure. From the figure we see that all our accepted parameter sets cluster to the high-$T$ end of the allowed region, with no models consistent with the SM-prediction $(S,T)=(0,0)$.  The offset along the $S$-direction is dominated by the technicolor degrees of freedom for which we use the perturbative estimate $S_{\rm{TC}}=1/(2\pi)$ \cite{Dietrich:2005jn}; the technicolor singlet new physics states contribute within a narrow range around this value. The $T$-direction on the other hand reflects the breaking of the custodial symmetry by the mass differences of the new physics states. The up and down techniquarks are taken mass degenerate leading to zero contribution to $T$, and the allowed values of $T$ then directly constrain the mass patterns of the weak doublet and triplet states.
 
%
%

%
\subsection{Direct dark matter detection limits}
%

For comparison of the model against DM searches, the relevant observables are the spin-dependent (SD) and spin-independent (SI) interaction cross sections with the constituents of atomic nucleus. In general our WIMPs have both SD and SI interactions, which follow from the WIMP couplings to the $Z$-boson given in Eq.~(\ref{Zcur}), and to the Higgs Boson given in Eq.~(\ref{eq:Higgscoup-mbasis}).  We will use the usual zero momentum transfer limit of the SI and SD WIMP nucleon cross sections, for which the experimental collaborations give constraints.  

The SI WIMP-nucleon cross section is given by
\begin{eqnarray}
\label{SI}
\sigma_{\rm{SI}}^{0} =  \frac{8 G_F^2}{ \pi}\frac{m_\N^2\mu_\N^{2}}{m_{H}^4} 
f_\N^2  |S_{ii}|^2\,,
\end{eqnarray}
where $m_\N$ and $m_{H}$  are the nucleon and the Higgs boson mass, respectively, and the mixing dependent $S_{ii}$-factor is given in Eq.~(\ref{eq_cfactors}). 
The effective Higgs-nucleon coupling factor 
\begin{equation}
f_\N \equiv \frac{1}{m_\N} \sum_{q} \langle \N|m_q \bar{q}q|\N \rangle \,,
\end{equation} 
describes the normalized total scalar quark current within the nucleon. The nucleon quark currents have been under intensive reseach over the last years by use of lattice and chiral perturbation theory methods and pion nucleon scattering~\cite{Bali:2011ks, Bali:2012qs,Alvarez-Ruso:2013fza,Oksuzian:2012rzb, Gong:2012nw, Freeman:2012ry,Engelhardt:2012gd, Junnarkar:2013ac, Young:2013nn, Jung:2012rz, Gong:2013vja,Gasser:1990ce, Gasser:1990ap, Borasoy:1996bx, Alarcon:2012kn, Alarcon:2012nr, Pavan:2001wz, Cheng:2012qr}. For a nice review see~\cite{Young:2013nn}. A statistical analysis based on these results was recently performed in ref.~\cite{Cline:2013gha} showing that $f_N$ is now fairly well determined; with a minor change from~\cite{Cline:2013gha} we use $\sigma_s = 40 \pm 10$ MeV as the input for strangeness matrix element (following~\cite{Junnarkar:2013ac}) and find $f_\N = 0.345 \pm 0.016$ at formal 1$\sigma$ level, in excellent agreement with~\cite{Cline:2013gha}. Thus the uncertainty in $f_N$ induces at most 20\% error (2$\sigma$) in the SI direct detection limits.

The SD WIMP-nucleon cross section~\cite{Angle:2008we,Primack:1988zm,Kainulainen:2009rb} in our model is 
\begin{equation}
\sigma_{\rm{SD}}^{0} = \frac{8 G_{F}^2}{\pi} \mu_\N^2 [a_p \langle S_p \rangle + a_n \langle S_n \rangle]^2 \frac{J+1}{J} \, |U_{1i}|^4 \,,
\label{SD}
\end{equation}
where $\mu_\N$ is the WIMP nucleon reduced mass, $a_{p,n}$ are the spin-dependent nucleon coupling factors and $\langle S_{p,n} \rangle$ are the proton and neutron spin expectation values. The total angular momentum of the nucleon is denoted by $J$, \ie $J=1/2$ for both protons and neutrons. We use the coupling values $a_p^2 = 0.46$ and $a_n^2 = 0.34$ for the pure proton and neutron cases, following from the EMC measurement values~\cite{Ellis:1994py}. Finally, the spin expectation values are $\langle S_p \rangle = 0.5$ and $\langle S_n \rangle = 0$ in the pure proton and $\langle S_n \rangle = 0.5$ and $\langle S_p \rangle = 0$ in the pure neutron case. 

\begin{figure}
\begin{center}
\includegraphics[width=0.7\textwidth]{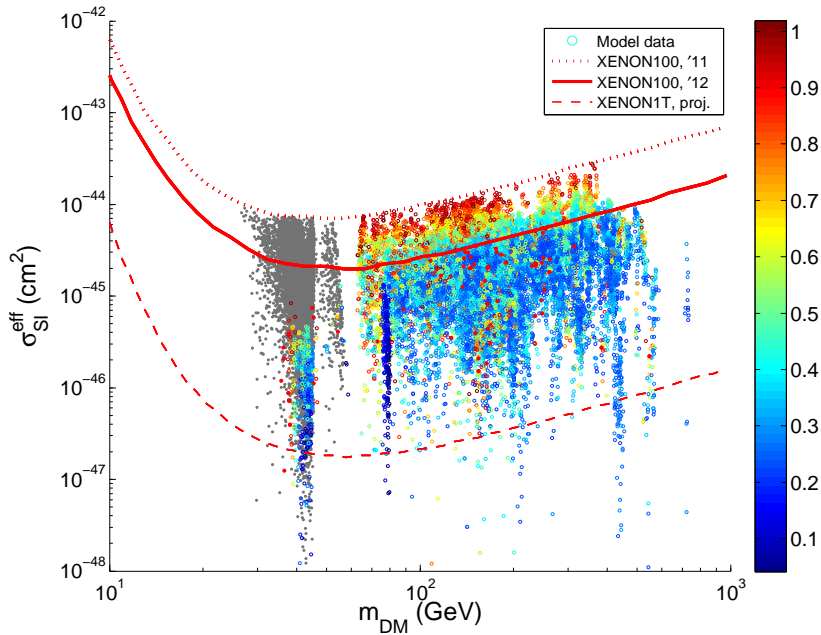} 
\caption{Model predictions for spin-independent WIMP-nucleon cross section for different WIMP masses.  Here we used the $2.6\sigma$ (S,T)-constraint, the Higgs width constraint $R_{\rm I} < 0.28$ and the spin independent XENON100 2011 limits to constrain the data. Coloring of data points encodes the fraction of relic density as indicated by the vertical bar on the right. Grey points are excluded by the invisible Higgs decay width constraint in Eq.~(\ref{eq:RIDM}). Also shown are the upper limits from the XENON 2011 exposure~\cite{Aprile:2012nq} (red dotted curve), the latest XENON100 2012 exposure~\cite{Aprile:2012nq} (solid red curve) and the projected XENON1T exposure (red dashed curve)~\cite{Aprile:2012zx}.}
\label{fig:SIa}
\end{center}
\end{figure}

\begin{figure}
\begin{center}
\includegraphics[width=0.7\textwidth]{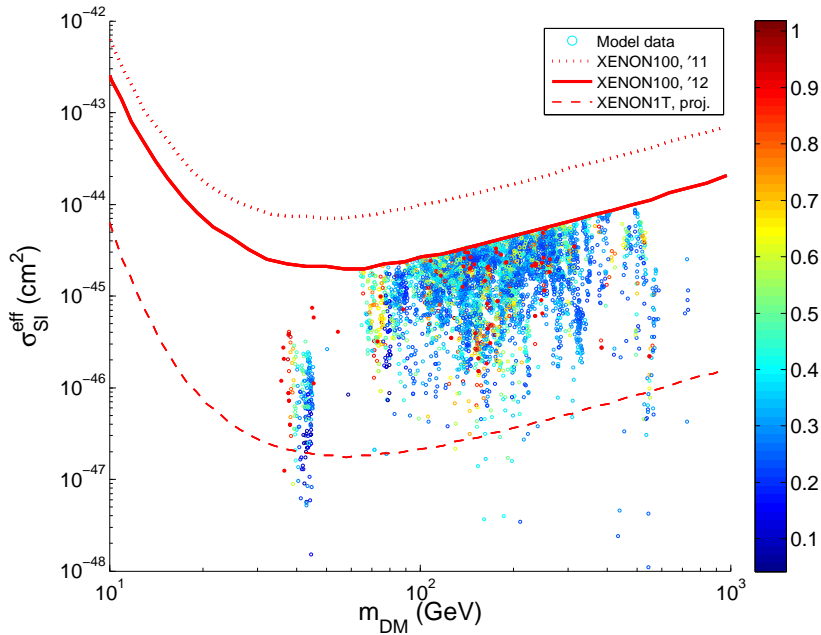}
\caption{Same as in figure \ref{fig:SIa}, except that more stringent constraints were used: $1.6\sigma$ for (S,T), $R_{\rm I} < 0.19$ and the spin independent XENON100 2013 and the spin dependent SuperK 2011 and Icecube 2012 limits. }
\label{fig:SIb}
\end{center}
\end{figure}

In this paper we are also interested in the cases where our WIMP forms but a part of the total amount of dark matter. In such a case the local WIMP density around Earth is not unambiguously defined. To break the ambiguity we assume that our WIMPs and the dominant DM component cluster in the same way, which implies that our WIMPs form a fraction 
\begin{equation}
f_{\rm rel} \equiv \Omega_\chi h^2/\ODM h^2\,,
\label{eq:frel}
\end{equation}
of the local DM-density. For the total density we used $\ODM h^2 = 0.1203$~\cite{Ade:2013zuv}. Second, to be conservative, we make no assumption on the observability of the dominant DM component, and consider the direct DM search constraints as applying on our WIMP candidate only. Now, the direct DM search constraints on $\sigma_{\rm{SI,SD}}^{0}$ are given under the assumption that $f_{\rm rel}=1$, and hence, under the above assumptions for subdominant WIMPs, these constraints should be applied on effective cross sections~\cite{Cline:2012hg,Cline:2013bln}
\begin{equation}
\sigma_{\rm{SI,SD}}^{\rm eff} \equiv f_{\rm rel}  \, \sigma_{\rm{SI,SD}}^{0} \,.
\end{equation}
We use these effective cross sections in our main results, which are summarized in figures~\ref{fig:SIa}, \ref{fig:SIb}, \ref{fig:SD} and \ref{fig:SDproton}. The advantage of this formulation is that one can immediately read how much a given direct search experiment needs to improve their sensitivity in order to rule out a given model parameter set.

%
%

Figures \ref{fig:SIa} and \ref{fig:SIb} display constraints for the SI WIMP-nucleon cross section with two different sets of laboratory and other direct and indirect detection  constraints as explained in figure captions. In figures \ref{fig:SD} and \ref{fig:SDproton} we show the constraints for the SD WIMP-neutron and for the SD WIMP-proton cross sections corresponding to the more stringent set of constraints as explained in the caption of figure \ref{fig:SIb}. In each plot we show the WIMP mass and the effective cross section pairs corresponding to parameter sets that satisfy all constraints in the MCMC-chain. Coloring of acceptable points displays the fraction of the relic density $f_{\rm rel}$ as indicated by the vertical bar to the right of each figure. The points colored grey would pass the direct DM detection constraints, but are excluded on the basis of the LHC Higgs data as will be explained in more detail in section \ref{Colliderconstraints}. We also show the current experimental limit from direct WIMP searches. Both for spin-independent and for spin-dependent WIMP-nucleus interactions in range $\mDM \gsim 10$ GeV best limits follow from XENON100 experiment~\cite{Aprile:2012nq,Aprile:2013doa}. These constraints are shown by red lines in figures~\ref{fig:SIa}, \ref{fig:SIb} and \ref{fig:SD}.\footnote{At smaller masses $\mDM \lsim 10$ GeV XENON limits are bettered by the CDMSII~\cite{Ahmed:2010wy} (SI) and by PICASSO~\cite{Archambault:2012pm} (SD WIMP-neutron). Also SIMPLE~\cite{Felizardo:2011uw} is more constraining than COUPP for $\mDM \lsim 20$ GeV, but these bounds are note relevant for us.} In the spin-dependent WIMP-proton channel, figure \ref{fig:SDproton}, we have shown (black line) the best direct search bound from the COUPP~\cite{Behnke:2012ys}-experiment.

Whether the SD- or SI-interactions are more constraining for the direct DM detection, depends on the specific values of the mixing matrix elements $U_{ij}$. For example in the case where WIMP has a substantial triplet-singlet $\beta\omega_0$-mixture the SI interaction is the dominant one, because in this case WIMP essentially only couples to Higgs boson. However, in case of a substantial doublet-singlet $\beta N_L$-mixture also SD interaction may be relevant because now WIMP couples also to the Z-boson. It turns out that the SI-interaction provide the strongest constraint for the most of our model parameter space. It is also possible that our WIMP has dominantly a pseudo-scalar ($P_{ii}$) coupling to the Higgs boson. In this case the WIMP-nucleus interaction is momentum transfer dependent, and therefore suppressed compared to the standard $Z$-mediated SD axial vector and Higgs mediated SI scalar interactions. Some of our parameter sets indeed fall into this category, in particular for $\mDM \approx 80$ GeV. These solutions can avoid being detected by any of the foreseeable direct and indirect DM searches.

%
%

Finally there has been reports on the observational side of DM signals in the low mass region:  DAMA/NaI/LIBRA~\cite{Bernabei:2000qi,Bernabei:2005hj,Bernabei:2008yi,Bernabei:2010mq}, CoGeNT~\cite{Aalseth:2010vx,Aalseth:2011wp,Aalseth:2012if}, CRESST~\cite{Angloher:2011uu} and CDMSII~\cite{Agnese:2013rvf}. Some of these observations contradict other existing bounds, but the discrepancies can be alleviated \eg by isospin violation or a long range forces in DM-nucleus interactions. However, small masses are not favored in our model and we shall not discuss these signals here. 

%
\subsection{Direct collider constraints}
\label{Colliderconstraints}
%

\begin{figure}
\begin{center}
\includegraphics[width=0.7\textwidth]{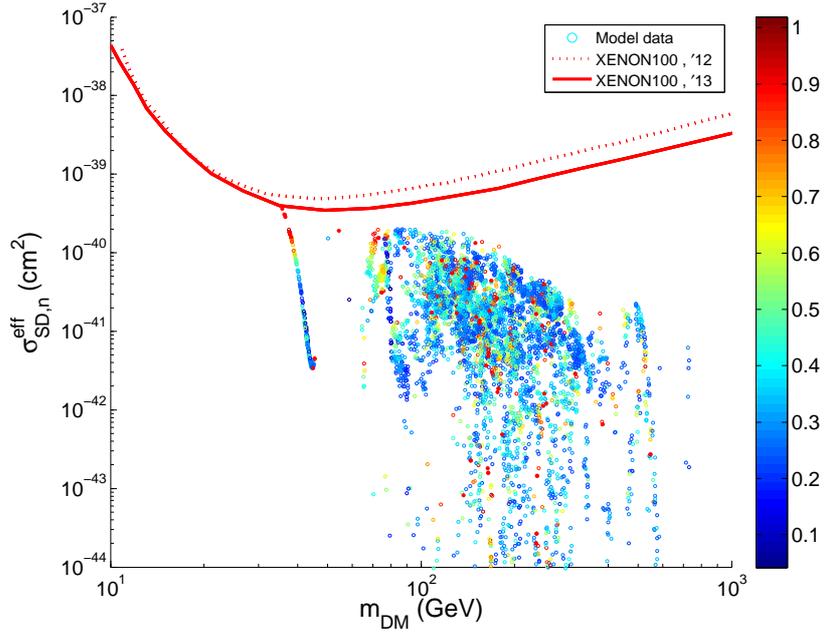}
\caption{Scatter of the accepted models as a function of the DM mass and the spin-dependent WIMP-neutron scattering cross section. Color mapping and the constraint selection is the same as in figure \ref{fig:SIb}. Red solid (dotted) curve is XENON100~\cite{Aprile:2013doa}  (XENON100~\cite{Aprile:2012nq} from~\cite{Garny:2012it})}
\label{fig:SD}
\end{center}
\end{figure}

{\em LEP limits.} In addition to constraining the oblique parameters, the LEP results impose two further constraints on our model: first, there is a bound coming from the $Z$-boson decay width measurement and second, LEP imposes lower bounds on the masses of new heavy charged particles. However, the latter bound, roughly equal to the maximum CM-frame collision energy in LEPII \ie $m \ge 104.5$, is by now superseded by LHC-constraints to be discussed below.  The $Z$-boson decay constraints are still relevant however. As $Z$-width into invisible channels is already saturated by light SM neutrinos, new physics contributions to the $Z$-width are highly constrained. This constraint is usually expressed in terms of a number of light neutrino species and the present experimental value is~\cite{Beringer:1900zz}:
\begin{equation}
N_{\nu} =\frac{\Gamma(Z\rightarrow{\rm{inv.}})}{\Gamma(Z\rightarrow\bar{\nu}\nu)}= 2.984 \pm 0.008 \,.
\end{equation}
The best fit value is already 2$\sigma$ below the SM prediction and any new physics would increase the tension. We choose to constrain our model by allowing at most one standard deviation from new physics, which in our case implies a bound
\begin{equation}
|U_{1i}|^4 \,\Big(1 - \frac{4 m^2_i}{m^2_Z}\Big)^{3/2} < 0.008 \,.
\end{equation}
This limit essentially rules out any light WIMPs with $\mDM < m_Z/2$ if the WIMP has a significant $N_L$ component. From figures \ref{fig:SIa},\ref{fig:SIb} and \ref{fig:SD} we see that there is a narrow set of allowed points clustering just below the limit $\mDM=m_Z/2$. However if the WIMP is an almost pure $\beta\omega_0$-mixture, the $Z$-decay width does not impose a strong constraint because $\omega_0$ does not couple to the $Z$-boson (in this limit $U_{12}\sim U_{13}\sim 0$). Instead, in this case a stringent constraint is provided due to the absence of invisible decay channel of the Higgs as will be discussed in the next subsection. The points falling to this category in our analysis are shown as grey points in figure \ref{fig:SIa}. \\

\begin{figure}
\begin{center}
\includegraphics[width=0.7\textwidth]{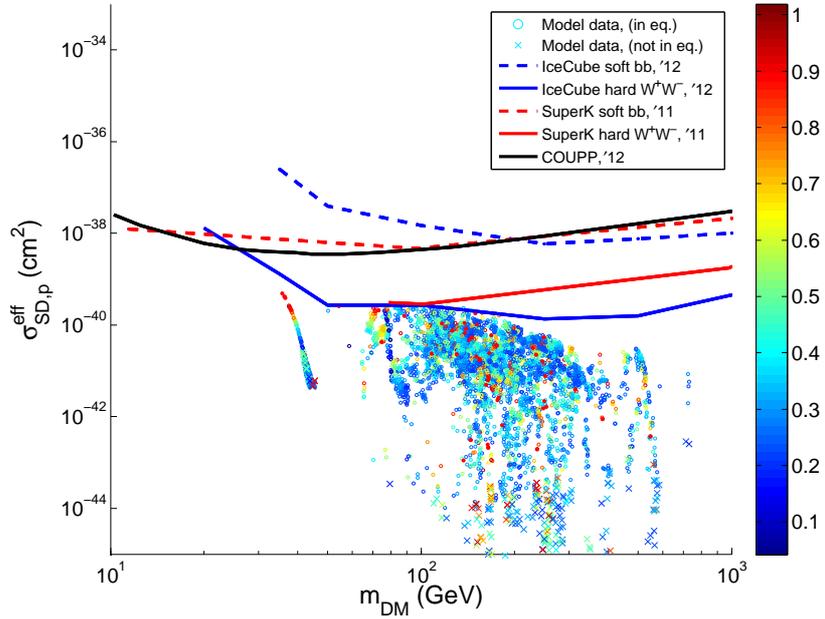}
\caption{Model prediction for spin-dependent WIMP-proton cross section for different WIMP masses. Color mapping and the constraint selection is the same as in figures \ref{fig:SIb} and \ref{fig:SD}. }
\label{fig:SDproton}
\end{center}
\end{figure}

\noindent {\em LHC limits.} LHC already provides strong limits on new charged particles. In our model these bounds affect the two charged Dirac particles: the fourth heavy electron $E$, and the charged adjoint state $\omega^{\pm}_D$. Although the LHC limits are not as straightforward to implement as those from LEP, they should in all cases be much stronger than the corresponding LEP limits. In our analysis we have used conservative bounds 
\begin{equation}
m_E, \, m_{\omega_D} > 500 {\rm GeV}.
\label{eq:boundonchargeds}
\end{equation}
The heavy electron,  $m_E\sim 1 $ TeV,  should easily avoid these limits and, moreover, $m_E$ is not directly related to the rest of the model parameters. Similarly to what was observed in ref.~\cite{Kainulainen:2009rb} however, $m_E$ in fact is correlated with other mass parameters through the oblique constraints. Yet, this correlation turns out to be fairly weak and so the current lower bound on $m_E$ is not very essential. The situation is somewhat different for the $\omega^{\pm}_D$ state, which belongs to the $\omega$-triplet and whose mass therefore appears also as one of the entries of the WIMP mass matrix: $m_{\omega_D} \equiv M_{\omega\omega}$. Thus the lower bound on $m_{\omega_D}$ is simultaneously a direct constraint on the WIMP mass matrix. All colored points in our figures fulfill bound~(\ref{eq:boundonchargeds}). Since these limits may be overly conservative (see \eg\cite{Aad:2011cwa}), we considered also a more relaxed bound $m_{\omega_{D,E}} > 200$ GeV. However, all viable parameter sets gained this way turn out to be excluded by the LHC constraint on invisible Higgs decay width. Indeed, if WIMP was lighter than $m_H/2$, then Higgs would decay to a pair of WIMPs with a rate:
\begin{equation}
 \Gamma_{\rm {H,DM}}  =  \frac{G_{F} m_H}{2 \sqrt{2} \pi} \Big(
 \,  |S_{ii}|^2 \beta_i^3 + |P_{ii}|^2 \beta_i  \, \Big) \,,
 \label{higgswidth}
\end{equation}
where $\beta_i \equiv (1- 4m_i^2/m^2_H)^{1/2}$ and the index $i$ refers to the WIMP as the lightest of the mass eigenstates. The invisible Higgs branching fraction would then be\footnote{The masses of the other new states i.e. $m_j, m_k, m_E $ and $ m_{\omega_D}$ are always $ > m_H/2$.}
\begin{equation}
R_{\rm I} =  \frac{\Gamma_{\rm {H,DM}}}{\Gamma_{\rm {H,DM}} + \Gamma_{\rm {SM,tot}}} \,,
\label{eq:RIDM}
\end{equation}
where $\Gamma_{\rm {SM,tot}}$ is the total Higgs decay width in the SM. The total invisible branching fraction is constrained to be~\cite{Giardino:2012dp,Dobrescu:2012td,Giardino:2013bma}

\begin{equation}
  R_{\rm {I}} \lsim 0.19 \;(0.28) \,.
\label{eq:stonginvicible}
\end{equation}
The more stringent of these bounds assumes SM-like Higgs-gauge field and Higgs-fermion couplings. This bound is relaxed if one accounts for uncertainties in the Higgs couplings in the effective low energy limit of the MWT model and the weaker constraint corresponds to the case where Higgs and SM gauge fields may have non-SM-like couplings to photons and gluons; see \cite{Giardino:2013bma} for details.\footnote{This scenario corresponds to how the extra EW triplet and doublet matter fields in our model would contribute to the experimental decay channels relevant for the LHC data.}
We explored the effect of both constraints. Note that changing Higgs-gauge couplings does not affect the WIMP mixing structure and hence the relic density in the mass range $m_{\rm \scriptscriptstyle DM} < 62.5$ GeV, where WIMPs dominantly annihilate to light fermions. Therefore there is no correlation between the two different treatments of the Higgs width bound and the relic density calculation. All colored points in figures (\ref{fig:SIa}-\ref{fig:SDproton}) represent parameter sets that fulfill the criteria~(\ref{eq:stonginvicible}). However, to illustrate the effect of the Higgs width constraint, we show also the sets excluded by~(\ref{eq:stonginvicible}) in gray.  \\
%
%
%
%
%

\noindent {\em Monojets.} Finally, we will comment about limits set for DM models using mono-jet/gauge boson data. Mono-jet searches in CDF~\cite{Aaltonen:2012jb}, ATLAS~\cite{ATLAS:2012ky, Rajaraman:2011wf} and CMS~\cite{Chatrchyan:2012me}, mono-photon searches in ATLAS~\cite{Aad:2012fw} and CMS~\cite{Chatrchyan:2012tea} and mono-W/lepton searches in~\cite{Bai:2012xg} and mono-Z searches in~\cite{Carpenter:2012rg}, all use effective field theory approach which assumes a heavy mediator linking the dark matter and the SM sectors. Therefore these results are not directly applicable to the case where mediators are light and can be produced on-shell in the experiment. In our case the mediators the Higgs- and the $Z$-boson are indeed light in this sense. Analyses appropriate also for light mediators were performed in refs.~\cite{Fox:2011fx} and~\cite{Fox:2011pm}, for the LEP mono-photon data and the LHC mono-photon and mono-jet data, respectively. (See also ref.~\cite{Fox:2012ru}). Adapting the results of~\cite{Fox:2011fx} and~\cite{Fox:2011pm} to our case, we conclude that no limits that would be comparable to the other constraints already discussed here arise from these data.

%
\subsection{Indirect dark matter detection limits}
%

If WIMPs interact with ordinary matter, they can accumulate in the cores of the Sun and the Earth. As these trapped WIMPs annihilate, they may produce neutrinos that can propagate to earth and produce muons that can be observed with the neutrino telescopes. If the WIMP capture and annihilation rates are in equilibrium, one can relate and constrain the WIMP annihilation and the WIMP-nucleus scattering cross sections through measurement of the muon flux induced by neutrinos originating from WIMP annihilations. So far no anomalous signal has been observed, and only constraints for the DM models can be set from the data. Most important indirect limits come from WIMP annihilations in Sun, and because the Sun is mostly made of hydrogen, the strongest constraints  are found on the SD WIMP-proton cross section.

If the rates are not in equilibrium, then the direct correlation between muon fluxes and WIMP-nucleus cross section is lost, and a more detailed analysis is needed to find constraints on the model. To evaluate if the equilibrium takes place in our model, we use Eqs. (1-7) from Ref.~\cite{Hooper:2003ui} (based on Refs.~\cite{Gould:1991hxnew,Jungman:1995df,Gould:1987ir}). We assume that WIMPs are in thermal equilibrium at solar core at temperature $T=T_{\odot} \approx 1.3$ keV. We denote by  $\tau = 1/ \sqrt{C_{\odot} A_{\odot}}$ the time scale during which the equilibrium between the WIMP capture rate $C_{\odot}$ and annihilation rate $\propto A_{\odot}$ can be achieved. Then the equilibrium is achieved with respect to the age of the solar system, $t_{\odot} = 4.5 \times 10^9$ provided $t_{\odot} / \tau \gg 1$. In figure \ref{fig:SDproton} we indicate by circles (crosses) the parameter sets for which $t_{\odot} / \tau > 1$ ($t_{\odot} / \tau < 1$). For more details, see \eg\cite{Hooper:2003ui,Wikstrom:2009kw}. 

For comparison with the data we assume that WIMPs annihilate with 100 $\%$ branching to $W^+ W^-$-pairs if $\mDM > m_W$ or to $\bar{\tau} \tau$-pairs if $\mDM < m_W$ (hard channel), or to $\bar{b}b$-pairs (soft channel). Neither of these assumptions holds for cases where our WIMPs annihilate dominantly to $Z Z, Zh$ or $\bar{t} t$, but we use this approach for simplicity, with the understanding that the resulting bounds may be too stringent.  Under these assumptions the strongest limits, especially for heavy WIMP masses, come from IceCube 2011~\cite{IceCube:2011aj} and 2012~\cite{Aartsen:2012kia} data in the hard $W^+W^-$channel and SuperKamiokande~\cite{Tanaka:2011uf} data, also in the hard $W^+W^-$channel. We show these limits in figure \ref{fig:SDproton} by solid red and blue curves, respectively. We also show the much weaker bounds arising from the assumption of the soft channel dominance in WIMP-annihlation by dashed red and blue curves. We expect the true bounds for our model space to lie somewhere between these two sets of curves, but closer to the hard channel bounds. Finally, we note that the current constraints derived for SI WIMP-nucleon interactions from indirect searches are not as stringent as the limits following from direct DM detection (\eg XENON100), and are not shown in the figures. 

WIMP annihilations can also take place other locations, such as in the Galaxy center, Galaxy halo or within smaller objects within the Galaxy, or in nearby dwarf galaxies.  The IceCube limits for these signals \cite{Abbasi:2012ws, Lee:2012pz,IceCube:2011ae} are not yet stringent enough. More stringent constraints in these cases are from FERMI-LAT gamma-ray data \cite{Ackermann:2012rg,Ackermann:2012qk,Pfrommer:2012mm,Ackermann:2011wa,GeringerSameth:2012sr}. However, these analyses assume that WIMPs annihilate in s-wave, while in our model, especially in WIMP mass range $\mDM \lsim 60$ GeV, the Majorana WIMPs tend to annihilate via Z-boson to light fermions, which proceeds in p-wave and is thus velocity suppressed. This makes the potential gamma signal weaker and comparing our model against these FERMI-LAT results is not straightforward. We leave the detailed study for future work, along with the analysis of possible future limits from the proposed CTA-observatory~\cite{Doro:2012xx,Wood:2013taa}.

Let us add a final comment that there are also indications of a 130 GeV DM signal in FERMI-LAT data (see \eg\cite{Bringmann:2012vr} and~\cite{Weniger:2012tx}). However, whether this is a true DM signal or not is still unclear. Other possible explanations include pulsar wind~\cite{Aharonian:2012cs} and instrumental systematics~\cite{Finkbeiner:2012ez},~\cite{Hektor:2012ev}. In light of the relative uncertainty of these signals, we do not attempt to explain them here.

%
\subsection{Discussion}
%

By far the most stringent dark matter detection limits on our model come from the XENON 2012 direct search bound on spin-independent WIMP-nucleon scatterings. Only in a domain around $\mDM \approx 100$ GeV the indirect constraints from IceCube and Super Kamiokande on the spin-dependent WIMP-proton interactions may provide a comparable bound, under certain simplifying assumptions on the WIMP annihilation channel. In particular, one can see from figure \ref{fig:SD} that all accepted models fall well below the current SD WIMP-neutron cross section bound and the situation is the same for the SD WIMP-proton interactions for $\mDM \gsim 200$ GeV, as seen in figure \ref{fig:SDproton}. For $\mDM < m_H/2$ the invisible Higgs width constraint provides a strong constraint, as we demonstrated in figure \ref{fig:SIa}. The other collider constraints, from precision electroweak data and the cuts on the masses of new charged fermions also provide important constraints. The precision electroweak data in particular has a potential to completely rule out our model if the limits on $S$-parameter improve to a level $S \lsim 0.1$, as is clearly evident from figure \ref{fig:STplot}. However, as seen by comparing figures \ref{fig:SIa} and \ref{fig:SIb} with different sets of collider bounds, these constraints do not single out particular areas in the ($\sigma,\mDM$)-planes. Barring significant improvement in the precision data the best hope of ruling out or verifying our model is improvement in the sensitivity of direct search SI WIMP-nucleon scattering, where one or two orders of magnitude improvement in the current sensitivity would be enough to rule out significant part of the remaining parameter space. Interestingly such  improvements are indeed expected in near future with LUX~\cite{Akerib:2012ak} and  XENON1T~\cite{Aprile:2012zx} experiments. Reaching the same sensitivity in the SD-channels would require 4-6 orders of magnitude improvement in detection efficiency, which is not likely to be achieved in near future. However, if one was lucky, a simultaneous observation of WIMP-events in all above channels could be used to accurately measure the different mixing parameters in our model.

%
\begin{figure}
\begin{center}
\includegraphics[width=0.7\textwidth]{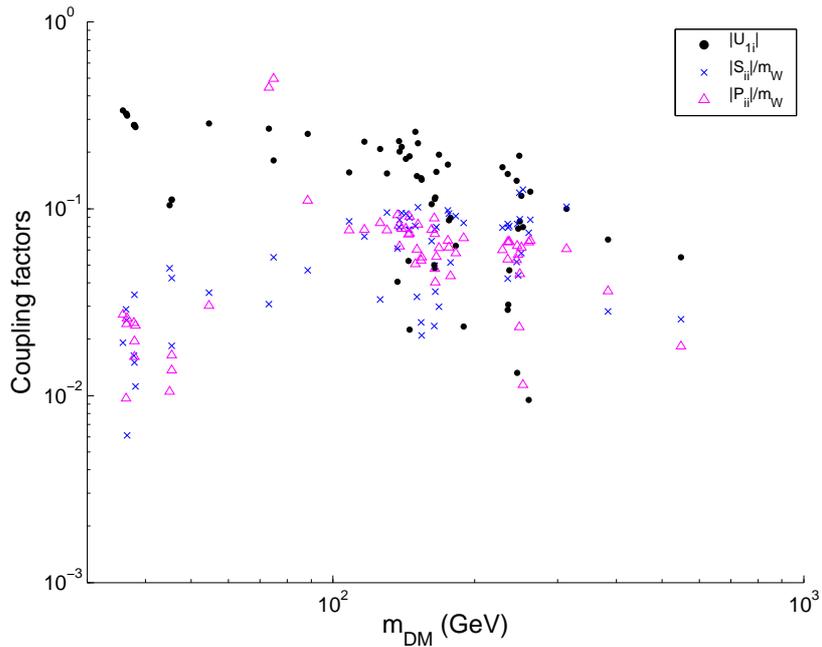}
\caption{Shown is the distribution of couplings $|U_{1i}|$ (circles), $|S_{ii}|/m_W$ (crosses) and $|P_{ii}|/m_W$ (triangles) for the subset of model parameter sets that provide a complete DM solution. Each parameter set is represented by a triplet of vertically aligned symbols.}
\label{fig:couplingplot}
\end{center}
\end{figure}
%

Let us stress that while we have chosen to display also parameter sets that would provide a subdominant dark matter component, our model can provide a complete solution to the dark matter problem. Indeed, our displayed data already contains a number of parameter sets that satisfy the current bounds on DM density, {\em i.e.} provide $\Omega_\chi h^2 =  0.1199 \pm 0.0027$~\cite{Ade:2013zuv}. Many more such models could be found simply by running longer MCMC chains. In figure \ref{fig:couplingplot} we show the distribution of the WIMP-$Z$-boson and scalar and pseudo-scalar WIMP-Higgs couplings for these sets. First, we see that all models have relatively small couplings, typically $|U_{1i}|\sim 0.1$ and $|S_{ii}|,\,|P_{ii}|\sim (0.01-0.1)m_W$. First, this shows that our WIMP is almost always mostly a singlet state. Second, the hierarchy between the $Z$- and Higgs couplings reflects the dominance of the SI WIMP-nucleon detection constraint; the relic density is mainly set by the $Z$-boson coupling, because Higgs-couplings are more strongly constrained. These features persist also in the full data-set including sub-dominant DM parameter sets. Finally, we note that pseudo-scalar couplings $|P_{ii}|/m_W$ (shown by triangles) are typically slightly larger than the scalar couplings $|S_{ii}|/m_W$ (shown by crosses). This is also expected, because the pseudo-scalar Higgs coupling to nucleons is further suppressed by velocity factors. Note in particular the two models at $\mDM \approx 80$ GeV in figure \ref{fig:couplingplot}, where the pseudo-scalar coupling is even larger than the $Z$-boson coupling; these are examples of models where DM can escape detection due to the velocity dependence of the direct search cross section\footnote{Let us note that one should use different nucleon matrix elements for the scalar and pseudo-scalar interactions, and the difference is very large for the WIMP-proton interaction (see for example~\cite{Cheng:2012qr}). We did not make this distinction because the velocity suppression in the pseudo-scalar interactions makes them less important. However, this approximation should be reconsidered if one had to resort more into use of parameter sets with a large hierarchy $|P_{ii}| \gg |S_{ii}|$ in future.}.

%
\begin{figure}
\begin{center}
\includegraphics[width=0.7\textwidth]{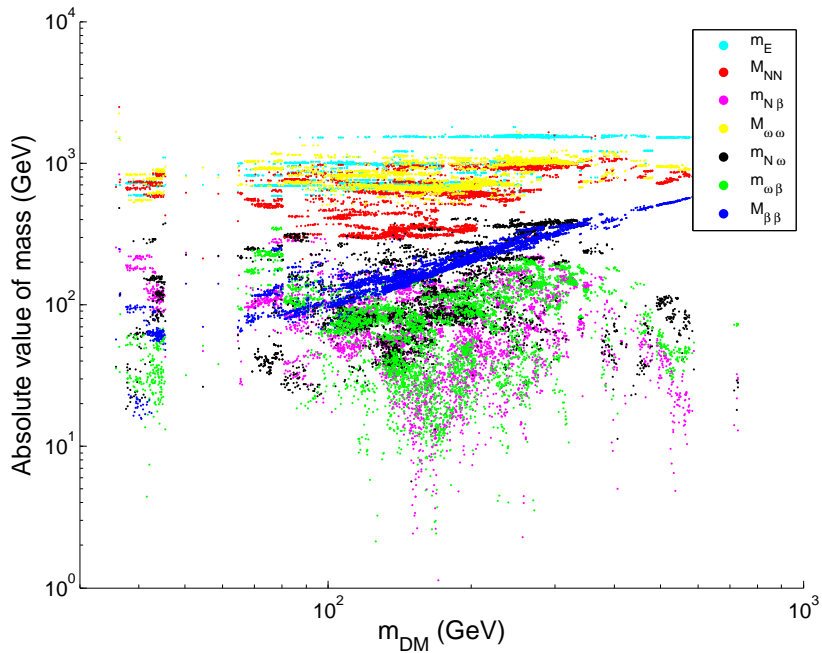}
\caption{Shown are the distribution of the elements of the input mass matrix (\ref{eq:massmatrix}) as a function of the WIMP mass for the data set corresponding to figures \ref{fig:SIb}, \ref{fig:SD} and \ref{fig:SDproton}.}
\label{fig:massplot}
\end{center}
\end{figure}
%

Let us finally study how the parameters of our input mass matrix (\ref{Adhiggsmass2}) and the charged lepton mass $m_E$ are distributed in the accepted models. In figure \ref{fig:massplot} we show the mass matrix element distributions as a function of the DM mass.  First, we see that the charged lepton mass terms $m_E$ and $\Mww$ have to be quite large and take values within ranges $0.8\lsim m_E/{\rm TeV} \lsim 1.2$ and  $0.8 \lsim \Mww/{\rm TeV} \lsim 1$, respectively. This is mainly due to the need to avoid direct detection in the collider experiments. Second, the Majorana mass $\MNN$ which, together with $\Mww$ mainly sets the heavier neutral mass eigenvalues tends also to be rather large $0.3\lsim \MNN/{\rm TeV} \lsim 1$. 
This is due the need to suppress the contributions of the other heavy states to the precision electroweak observables, in particular to $T$ parameter which is sensitive to the mass splitting within the fourth generation lepton doublet. The positive slope of the narrow band occupied by the mass parameter $\Mbb$ as a function of $\mDM$ illustrates the fact that DM tends to be dominantly a singlet ``bino"-like state. The Dirac masses $\mNb$ and $\mwb$, restricted to be roughly below $300$ GeV, are small due to the same reason; these parameters control the size of the mixing angles of the WIMP with the active states, which need to be small. Finally, the Dirac mass parameter $\mNw$, which controls the doublet-adjoint mixing of the WIMP is less constrained, but in general smaller than the diagonal Majorana masses.

%
\section{Conclusions}
%

We have described a simple extension of the SM. Our model leads to the unification of the SM coupling constants, breaks EW symmetry dynamically by a new strongly coupled sector and features a novel sector of neutral states leading to a WIMP candidate. The model studied here combines and extends earlier work carried out in refs.~\cite{Kainulainen:2010pk,Kainulainen:2009rb}. We have considered the general mass and mixing patterns in the three-dimensional space of neutral states, evaluated the resulting WIMP relic density and WIMP nucleon cross sections, and contrasted the results against existing direct DM detection and collider data. The parameter space scan was carried out with a MCMC method. We showed that while current and future experiments impose severe constraints on the model, there is a large portion of the parameter space where the the model can provide a full or a subdominant dark matter component. Most of the parameter space is testable in the future direct DM search experiments using the spin-independent WIMP-nucleon channels. Moreover, all our accepted parameter sets predict a positive $S$-parameter $S\lsim 0.1$ whereby the model can be potentially ruled out by improved precision electroweak data. In the case of a positive detection in all different direct and/or indirect DM-search channels, one can hope to infer the underlying mixing parameters in addition to the DM-mass.
	 
Among future directions of research would be a more accurate analysis of the indirect search bounds from neutrino telescopes and computing the FERMI-LAT and CTA-bounds from the gamma-ray fluxes. These are likely to remain subdominant against the direct DM searches using spin-independent cross sections, but they would be relevant in the case of a positive detection. Also, it will be interesting to study the model with complex mass parameters and CP-violation, in the context of electroweak baryogenesis. Further analyses of the model could also be carried out taking into account the strongly coupled states beyond the scalar meson sector, and their effect on both the collider phenomenology and dark matter.

\acknowledgments
The CP$^3$-Origins centre is partially funded by the Danish National Research Foundation, grant number DNRF90. 

 
\appendix
\section{Oblique parameters}
\label{sec:appendix}

Here we give the formulas for the oblique parameters $S$ and $T$ in our model. We base our analysis on~\cite{Antipin:2009ks}, which partially relies on works~\cite{Dietrich:2005jn,Dietrich:2005wk,Kainulainen:2006wq,Gates:1991uu,Holdom:1996bn,Kniehl:1992ez}. If compared to~\cite{Antipin:2009ks}, here the new contributions to the $S$ and $T$ follow from the $\omega$-triplet state and from the new WIMP mixings. The $S$ parameter is defined as
\begin{equation}  
S \equiv - 8 \pi  \frac{{\rm d} \Pi_{3 Y} (q^2) }{{\rm d q}^2} |_{q^2 = 0} \approx  -  \frac{8 \pi  }{m_Z^2} (\Pi_{3 Y} (m_Z^2)-\Pi_{3 Y} (0)),
\label{Sparam}
\end{equation}
and the $T$ parameter
\begin{equation}  
T \equiv    \frac{4 \pi  }{\sin^2 \theta_W \cos^2 \theta_W m_Z^2} (\Pi_{11} (0)-\Pi_{33} (0)),
\label{Sparam}
\end{equation}
where $\theta_W$ is the Weinberg angle and the subscripts $Y$, 3 and 1 refer to the hypercharge, and to the weak isospin components respectively. The contributions to gauge boson self energies in our model are   
\begin{eqnarray}  
\Pi_{3 Y} (q^2) &=& \sum_i  \frac{1}{2}|U_{ii}|^4 ( \Pi_{L L}(m_i^2,m_i^2,q^2)
                                                   -\Pi_{L R}(m_i^2,m_i^2,q^2) ) 
\nonumber \\
                &+& \sum_{j>i} \Big[  (|A_{ji}|^2 - |V_{ji}|^2) \Pi_{L L}(m_i^2,m_j^2,q^2) 
                                    - (|A_{ji}|^2 + |V_{ji}|^2) \Pi_{L R}(m_i^2,m_j^2,q^2) \Big]  \nonumber \\
                &-&  \frac{1}{2} \Pi_{L L}(m_E^2,m_E^2,q^2)-\Pi_{L R}(m_E^2,m_E^2,q^2)  ,
\label{Pi3Y}
\end{eqnarray}
and
\begin{eqnarray}  
\Pi_{1 1} (0) - \Pi_{3 3} (0) 
      &=&  \sum_i \Big[ \frac{1}{2}|U_{ii}|^2  \Pi_{L L}(m_i^2,m_E^2,q^2)  \nonumber \\
      &+&  2 (V_i^2 + A_i^2) \Pi_{L L}(m_i^2,m_{\omega^\pm}^2,q^2) 
         + 2 (V_i^2 - A_i^2) \Pi_{L R}(m_i^2,m_{\omega^\pm}^2,q^2)  \nonumber  \\
      &-&  \frac{1}{4}|U_{ii}|^2 ( \Pi_{L L}(m_i^2,m_i^2,q^2)
                                  -\Pi_{L R}(m_i^2,m_i^2,q^2) ) \Big] \nonumber \\
      &+& \sum_{j>i} \Big[ \frac{1}{2} (A_{ji}^2 + V_{ji}^2) \Pi_{L L}(m_i^2,m_j^2,q^2)                  
                         - \frac{1}{2} (A_{ji}^2 - V_{ji}^2) \Pi_{L R}(m_i^2,m_j^2,q^2) \Big]  \nonumber \\
      &-& \Pi_{L L}(m_{\omega^\pm}^2,m_{\omega^\pm}^2,q^2)
         -\Pi_{L R}(m_{\omega^\pm}^2,m_{\omega^\pm}^2,q^2)  
\nonumber \\
      &-&  \frac{1}{4} \Pi_{L L}(m_E^2,m_E^2,q^2),
\label{Pi11}
\end{eqnarray}
where the mixing dependent coupling factors 
$A_{ji}, B_{ji}, A_{i}$, and $ B_{i}  $ are given in (\ref{AB}), and the vacuum polarizations of left and right handed currents are 
\begin{eqnarray}  
  \Pi_{L L}(m_1^2,m_2^2,q^2) &=& -\frac{4}{(4 \pi)^2} \int_0^1{\rm d}x
   \ln \left[ \frac{\mu^2} {M^2 - x(1-x) q^2} \right] \left( x(1-x) q^2 - \frac{1}{2} M^2\right) 
\\
  \Pi_{L R}(m_1^2,m_2^2,q^2) &=& -\frac{4}{(4 \pi)^2} \int_0^1 {\rm d} x 
   \ln \left[ \frac{\mu^2} {M^2 - x(1-x) q^2}  \right]  \frac{1}{2} m_1^2 m_2^2 , 
\label{PiLL}
\end{eqnarray}
with $M^2 = x m_1^2 + (1-x) m_2^2$. For the cutoff $\mu$ we have used 1.5 $\times \max \{m_i, m_j, m_k \}$ through out the analysis. Finally, the contributions from the Technicolor sector to oblique parameters $S = 1/(2 \pi)$ and $T=0$, that we used in our analysis, follow from the usual naive perturbative estimate. 

%
\bibliography{KTV4_KT_3_7.bib}
%

\end{document}